\documentclass{aa}
\usepackage{graphicx}

\begin{document}

        \title{XMM-Newton and INTEGRAL observations of the black hole candidate XTE~J1817-330
	\thanks{Based on observations obtained with XMM-Newton, an ESA science mission with 
	instruments and contributions directly funded by ESA Member States and NASA; and with INTEGRAL, 
	an ESA project with instruments and science data center funded by ESA member states 
	(especially the PI countries: Denmark, France, Germany, Italy, Switzerland, Spain), 
	Czech Republic and Poland, and with the participation of Russia and the USA.}}

        \author{G.~Sala, J.~Greiner, M.~Ajello, E.~Bottacini, F.~Haberl}

	\institute{Max-Planck-Institut f\"ur extraterrestrische Physik, Postfach 1312, D-85741 Garching, Germany\\
			\email{gsala@mpe.mpg.de}}

        \offprints{G. Sala}

        \date{Received ... /accepted ...}

 
  \abstract
{}
    {The galactic black hole candidate XTE~J1817-330 was discovered in outburst by RXTE in January 2006. 
We present here the results of an XMM-Newton Target of opportunity observation (TOO), 
performed on 13 March 2006 (44 days after the maximum), 
and an INTEGRAL observation performed on 15-18 February 2006 (18 days after the maximum).}
    {The EPIC-pn camera on-board XMM-Newton was used in the fast read-out Burst mode to avoid photon pile-up, 
while the RGSs were used in Spectroscopy high count rate mode. 
We fit both the XMM-Newton and the INTEGRAL spectra  with a two-component model 
consisting of a thermal accretion disk (represented by \emph{diskbb} or \emph{diskpn} models) and
a comptonizing hot corona (represented by a power-law or the \emph{compTT} model).}
   {The soft X-ray spectrum is dominated by an accretion disk component, 
with a maximum temperature decreasing from 0.96$\pm0.04\,\mbox{keV}$ at the time
of the INTEGRAL observation to $0.70\pm0.01\,\mbox{keV}$ on 13~March.
The Optical Monitors on board INTEGRAL and XMM-Newton showed the source 
with magnitudes V: 11.3--11.4, U:15.0--15.1 and UVW1:14.7--14.8.
The soft X-ray spectrum, together with the optical and UV data, show a low hydrogen 
	column density towards the source, and 
	several absorption lines, most likely of interstellar origin, are detected in the RGS spectrum: 
	OIK$\alpha$, OIK$\beta$, OII, OIII and OVII, which trace both cold and 
	hot components of the ISM. The soft X-ray spectrum indicates the presence of a black hole, 
	with an estimate for the upper limit of the mass of 6.0$^{+4.0}_{-2.5}$M$_\odot$. }
{}



        \titlerunning{The black hole candidate XTE~J1817-330}
        \authorrunning{G. Sala~et~al.}
        \maketitle

%
%

\section{Introduction}
\label{intro}

X-ray binaries are the brightest X-ray sources in the sky. They are powered
by the accretion of material from the secondary star 
onto a compact object (neutron star or black hole). The spectral type of the 
companion determines the accretion regime and properties. In low mass X-ray binaries, 
with secondary stars of type later than A, the mass transfer occurs through Roche lobe
overflow and forms an X-ray emitting accretion disk.  
The optical emission is dominated by the X-ray heated companion,
the outer disc and/or reprocessed hard X-rays. 
In the case of high mass X-ray binaries, where the companion is an O or B 
star, the strong wind of the secondary star is intercepted and accreted by the compact object. 
The main X-ray source stems from the wind interaction. Though an accretion disk may also be present,
the secondary star is dominating the optical emission of the source.

The generally accepted picture for the X-ray emission of accreting black holes consists of 
an accretion disk, responsible for thermal black body emission in the X-ray band;
and a surrounding hot corona, origin site of non-thermal power-law emission, 
up to the energy range of gamma-ray telescopes, due to inverse comptonization 
of soft X-ray photons from the disk. The standard accretion disk model 
(Mitsuda et al. \cite{mit84}, \emph{diskbb} model in \emph{xspec}) 
consists of the sum of blackbodies from an accretion disk with a surface temperature 
that depends on the radius $R$ as $R^{-3/4}$. This asymptotic distribution is sufficiently 
accurate for $R\gg6 R_{\rm{g}}$. But since the model neglects the torque-free boundary condition, 
the temperature distribution is not accurate for the inner disk when it extends down to the 
innermost stable orbit. The \emph{diskpn} model in \emph{xspec} (Gierli\'nski~et~al. \cite{gie99})
includes corrections for the temperature distribution near the black hole 
by taking into account the torque-free inner-boundary condition.

At present, around 20 X-ray binaries contain a dynamically confirmed black hole, and 
around another 20 are the so called black-hole candidates (see Remillard \& McClintock \cite{rmc06} for 
a recent review). Seven of the 20 confirmed black holes, and 12 of the black-hole candidates
are transient sources with only one unique outburst observed. XTE~J1817-330 is the next one to be 
added to the list of transient black-hole candidates.

\begin{figure}
\centering
\resizebox{\hsize}{!}{\includegraphics{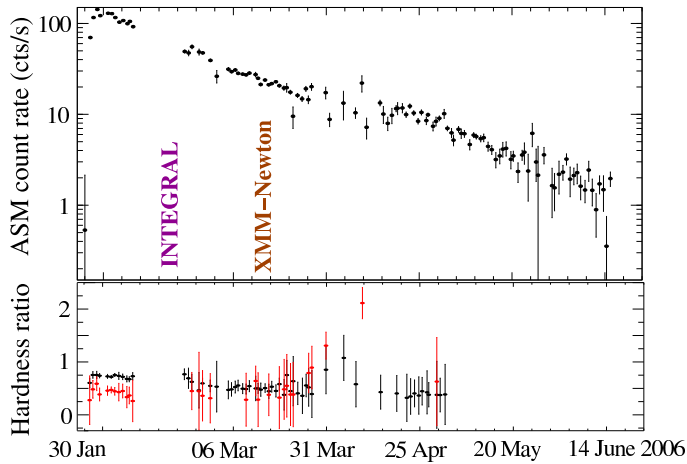}}
\caption{RXTE/ASM light curve of XTE J1817-330 in the 2-12 keV energy band (top panel),
and hardness ratios: HR1=(3-5~keV)/(1.5-3~keV) in black, and HR2=(5-12~keV)/(3-5~keV) in red (bottom).
The epochs of the XMM-Newton and INTEGRAL observations are indicated.}
\label{fig1}
\end{figure}

XTE~J1817-330 was discovered by the Rossi X-ray Timing Experiment (RXTE) on
26 January 2006 (Remillard~et~al.~\cite{rem06}) with a flux of 0.93($\pm0.03$)~Crab (2-12 keV) and 
a very soft spectrum, typical for black hole transients. The source reached
a maximum of $\sim1.9$~Crab on 28 January 2006, followed by an exponential decline in 
X-ray flux (Fig.~\ref{fig1}), with an e-folding time of 27 days.
Only five days after the discovery, the radio counter-part was identified, with a 
steep spectrum and no evidence for variability (Rupen et al.~\cite{rup06a}), 
and a second observation four days later showed a fading in 
flux of a factor $\sim$2.6 (Rupen et al.~\cite{rup06b}).
The NIR counterpart was discovered on 7 February with magnitude $K=15.0\pm0.4$ (D'Avanzo et al.~\cite{dav06}), 
and only four days later, the optical counterpart was identified (Torres et al.~\cite{tor06}), 
with $g'=14.93\pm0.05$.
The g'-K color supports a low reddening to the source. 
The non-detection of the source in previous images indicates a 
large outburst amplitude in the optical, suggesting a short orbital period (Torres et al.~\cite{tor06}).

\begin{table*}
\centering
\caption{\label{tab_bat} Swift/BAT observations (errors 90\% confidence).}
\begin{tabular}{c c c c c}
\hline\hline
\noalign{\smallskip}
Obs. ID & Observation start  & Observation end  & Flux (10--200~keV)  & Photon~index\\
	&	 	&		 & ($\times10^{-9}\mbox{erg}\,\mbox{cm}^{-2}\mbox{s}^{-1}$)	&	\\
\noalign{\smallskip}
\hline
\noalign{\smallskip}
00030359011 & 2006-01-31T22:15:43 & 2006-01-31T22:38:02 & $2.5\pm0.7$ & $2.2\pm0.4$ \\
00035348002 & 2006-02-03T03:21:26 & 2006-02-03T03:34:02 & $3.1\pm0.2$ & $2.7\pm0.5$ \\
	    & 2006-02-03T06:39:36 & 2006-02-03T06:49:18 & $2.8\pm0.8$ & $2.6\pm0.3$ \\
	    & 2006-02-03T17:48:27 & 2006-02-03T18:01:01 & $2.1\pm0.7$ & $2.7\pm0.6$ \\
            & 2006-02-03T21:01:26 & 2006-02-03T21:18:02 & $2.3\pm0.6$ & $2.9\pm0.2$ \\
00035339002 & 2006-02-15T14:13:03 & 2006-02-15T14:35:02 & $1.2\pm0.5$ & $2.1\pm0.5$ \\
00192152001 & 2006-02-24T14:46:31 & 2006-02-24T15:29:12 & $2.7\pm0.3$ & $2.5\pm0.2$ \\
	    & 2006-02-24T16:23:00 & 2006-02-24T17:05:38 & $3.1\pm0.3$ & $2.6\pm0.2$ \\
	    & 2006-02-24T17:59:26 & 2006-02-24T18:42:04	& $3.2\pm0.3$ & $2.5\pm0.2$ \\
	    & 2006-02-24T19:35:53 & 2006-02-24T20:15:21 & $2.8\pm0.3$ & $2.5\pm0.2$ \\
\hline
\noalign{\smallskip}
\end{tabular}
\end{table*}

We requested an XMM-Newton (0.4-10.0 keV) Target of Opportunity Observation (TOO) shortly after the discovery.
The observation could not be performed until the first visibility period of the source, 
which started on 13 March 2006. Here we report on the results of the XMM-Newton TOO, together with 
an INTEGRAL (20-150 keV) pointed observation performed on 15-18 February 2006 (Goldoni et al.~\cite{gol06}). 
In section \ref{sec_obs} we give the details of the observations. 
The results are described in section \ref{sec_ana} and discussed
in section \ref{sec_dis}.

\begin{figure}
\centering
\resizebox{\hsize}{!}{\includegraphics[angle=0]{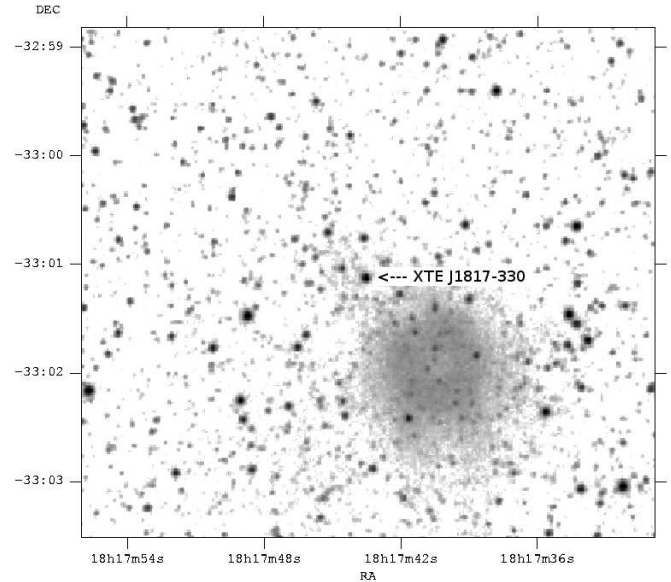}}
\caption{U band image of the XTE J1817-330 field of view obtained by the Optical Monitor on board XMM-Newton, 
adding the five individual 1.7~ks exposures. The irregular spot in the low-right part of the 
image is an instrument artifact caused by ``stray-light''.}
\label{om}
\end{figure}

\section{XMM-Newton and INTEGRAL observations}
\label{sec_obs}

We obtained a TOO observation with XMM-Newton (0.1-10.0~keV) 
on 13 March 2006 (obs. ID. 0311590501, 20~ks), when the source flux detected by 
the All Sky Monitor (ASM) on board RXTE had faded to $\sim$300~mCrab. 
The EPIC-pn camera was used in the fast read-out Burst mode (Kuster~et~al.~\cite{kus99}). 
The spectrometer RGS1 was operated in Spectroscopy High Count Rate mode, which reads the CCD nodes one by one, 
avoiding photon pile-up. RGS2 cannot be operated in this mode and was highly affected by pile-up. 
The two MOS cameras were switched off for telemetry limitations. 
The Optical Monitor (OM) on board XMM-Newton 
obtained simultaneous optical light-curves in the Fast mode, using filters U 
(first 10~ks, split in five 1.7~ks exposures) and UVW1 (10~ks, again split in five 1.7~ks exposures), 
and using simultaneously a fast mode window and an imaging window. 
 
XMM-Newton data were reduced with the XMM-Newton Science Analysis Software (SAS 7.0.0), and
XRONOS 5.21 and XSPEC 11.3 were used for timing and spectral analysis. 
The high flux of the source provides a spectrum with high statistics in the EPIC-pn camera.
After checking that adding the double events increases the 
total count rate only by 10--15\%, we decided to keep the best spectral resolution for the EPIC-pn by taking 
only single events (pattern=0) into account.
The source spectrum was extracted from a box including 10~RAWX columns at each side of 
the source position, and excluding events with RAWY~$\geq$~140 (to avoid direct illumination by the source) 
and RAWY~$\leq$~10. 
Some problems with the calibration of the Burst mode are affecting the soft 
part of the EPIC-pn spectrum, and cause extra residuals in this energy range, especially 
around the oxygen edge at 0.5~keV. We therefore consider events with
energies between 0.6 and 10.0~keV for spectral analysis. The exclusion of energies below 0.6~keV
does not represent an important loss in this case, since the lowest end of the 
X-ray band is simultaneously covered by the RGS, which provide good spectra at high resolution 
of this energy band.

The OM data have been included for joint spectral analysis with the X-ray data by using the 
canned response matrices provided at the XMM-Newton calibration portal (v1.0) 
in conjunction with two spectral data files containing the average count rates for each filter. 
The total count-rate error has been determined by adding the statistical filter 
measurement error given by the OM processing, in quadrature with the 
absolute error of the calibration of the response matrices, which is 10\%.

INTEGRAL observed XTE~J1817-330 in hard X-rays (20-150~keV) 
as a TOO on 15-18 February 2006 (Goldoni et al.~\cite{gol06}). 
90 science windows for a total amount of 193~ks of exposure time on the source were considered.
We used the INTEGRAL Off-line Scientific Analysis (OSA) version 5.1, and the
matrices available for standard software  
(\emph{isgr\_rmf\_grp\_0016.fits} and \emph{isgr\_arf\_rsp\_0014.fits}) 
for the spectral analysis.
Data screening was performed according to the median count rate with respect to each science window
and its distribution: science windows with rate larger than 10 standard deviations from the 
median count rate were excluded. We added a systematic error of 10\% to the JEM-X data.

XTE~J1817-330 was also detected with Swift/BAT.
We analyzed 7 months of BAT survey data (from January 2006 to July 2006)
looking for a detection of XTE~J1817-330. Data processing was
performed using standard Swift software available in the HEADAS 6.1
distribution and following the recipes of the BAT analysis threads
at the HEASARC web pages. Specifically, only data which satisfied the
following requirements were used:
(1) the star tracker is locked and the spacecraft is settled on the nominal
pointing direction;
(2) the spacecraft is outside the South Atlantic Anomaly and the total rate
of  the BAT array is less than 18000~s$^{-1}$ .
Data which fulfilled the above requirements were rebinned in
energy accordingly to the Gain/Offset map generated on board and were
used to generate a sky image. The sky images were searched for excesses
above the 6-sigma threshold.
XTE J1817-330 was detected 10 times above this threshold and for all
these observations we extracted a spectrum (Table~\ref{tab_bat}). The BAT XTE J1817-330
spectra were used to test the spectral variations between the XMM-Newton
and the INTEGRAL observations.

\begin{figure}
\centering
\resizebox{\hsize}{!}{\includegraphics[angle=0]{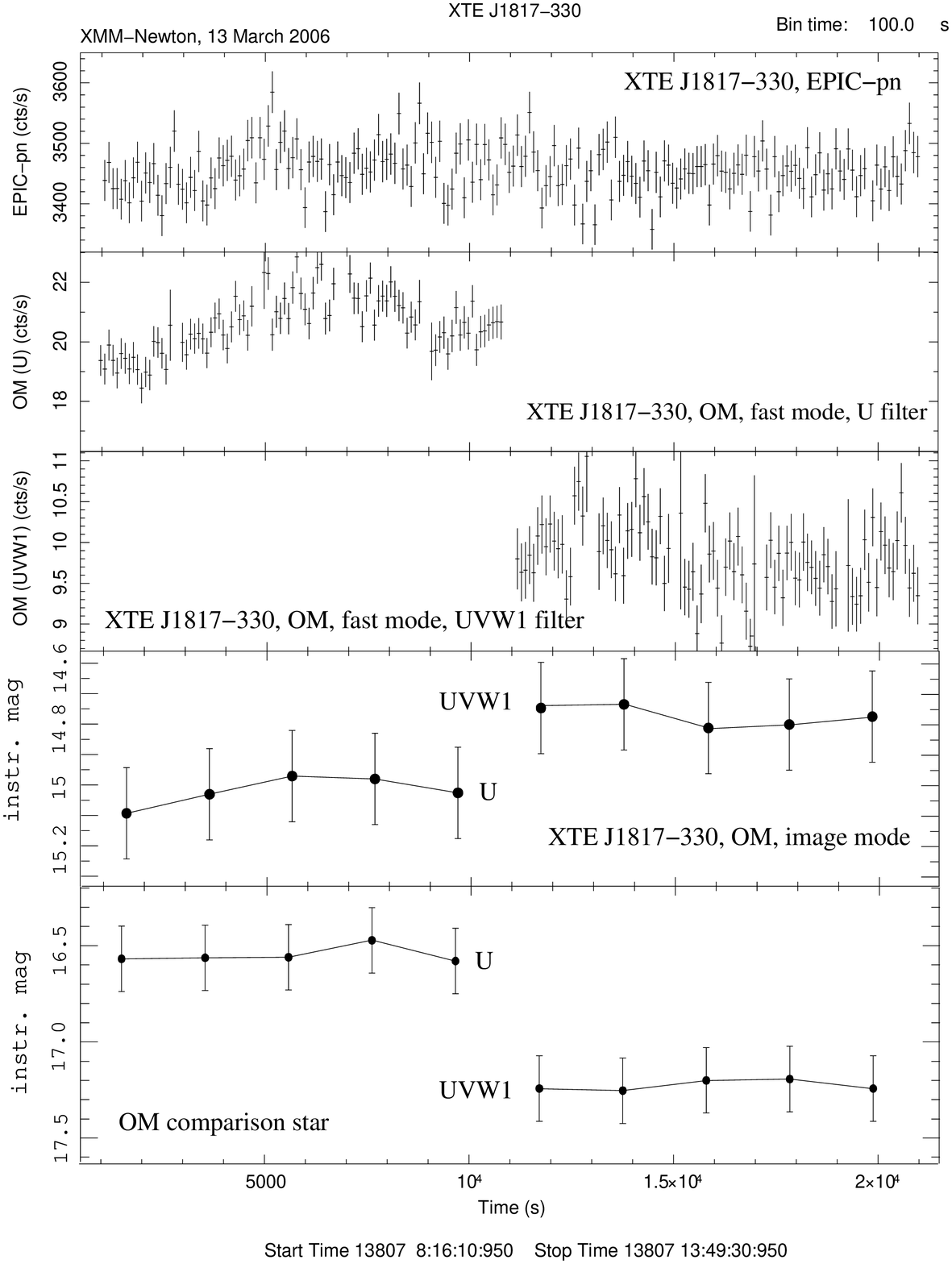}}
\caption{XTE J1817-330 X-ray, optical and UV light curves on 13 March 2006: 
count-rates from EPIC-pn (dead-time corrected, upper panel);
OM fast mode window exposures with U and UVW1 filters  (second and third panels);
instrumental magnitudes (in the XMM-Newton Optical Monitor AB system, 
$mAB=zero.point(filter)-2.5\log10(count.rate)$, see Chen~et~al.~\cite{che04} for details) from the OM image 
mode window exposures with U and UVW1 filters (fourth panel); 
and light curve of a comparison star in OM field of view (lower panel).}
\label{figlc}
\end{figure}

\section{Results \label{sec_ana}}

XTE~J1817-330 was clearly detected by all instruments on board XMM-Newton and INTEGRAL.
It provided a high statistics spectrum in the EPIC-pn camera, with a count-rate of 3390$\pm$60~cts/s.
The XMM-Newton Optical Monitor shows the source with 
standard AB magnitudes $U=15.95\pm0.01$ and $UVW1=16.12\pm0.01$ 
(magnitudes from the OM observation source list obtained by the SAS task \emph{omsrclistcomb})
(Fig.~\ref{om}).

The EPIC-pn light curve, together with the simultaneous U and UVW1 light curves 
obtained by the XMM-Newton OM, are shown in Fig.\ref{figlc}. 
The light curve of a comparison star in the OM field of view is also shown.
The EPIC-pn count-rate is corrected for dead time losses, which in Burst mode are 97\%.
The X-ray light curve is constant within 5\% during the whole observation
The simultaneous U light curve shows a slight modulation, rising during the first $\sim$6~ks
and decreasing afterwards. Although not so clearly, the later UVW1 light curve suggests a possible
continuation of this modulation. This might hint to an orbital modulation with a period of 6~hours.

XTE~J1817-330 is clearly detected in IBIS/ISGRI, with a significance of 246$\sigma$,
with a total of 2.4$\times10^6\mbox{cts}$ and a count rate of $18.3\pm0.1\,\mbox{cts s}^{-1}$.
The Optical Monitor on-board INTEGRAL shows XTE~J1817-330 with an instrumental
visual magnitude of V: 11.35$\pm$0.05.

\subsection{Spectral analysis}

We use a 2-component model consisting of a thermal accretion disk plus power-law to fit both
the XMM-Newton and INTEGRAL spectra of XTE~J1718-330. 
Since the RXTE/ASM hardness ratio of the source is
not changing substantially from the start of the outburst until the time of the XMM-Newton
observation, we checked whether the spectrum was the same in February (time of INTEGRAL observation)
as in March (XMM-Newton observation). For that purpose we have analyzed 
the spectra of 10 Swift/BAT XTE~J1817-330 detections 
in February--March 2005, five of them occurring between the INTEGRAL and the XMM-Newton observations 
(Table~\ref{tab_bat}).
The BAT spectra in the 20--60~keV range during this period are indeed consistent, within the errors, 
with a power-law of photon index 2.5--3.0. This range is however too wide as to assume a 
constant spectrum, so we fit separately INTEGRAL and XMM-Newton data.

\begin{figure}
\centering
\resizebox{\hsize}{!}{\includegraphics[angle=-90]{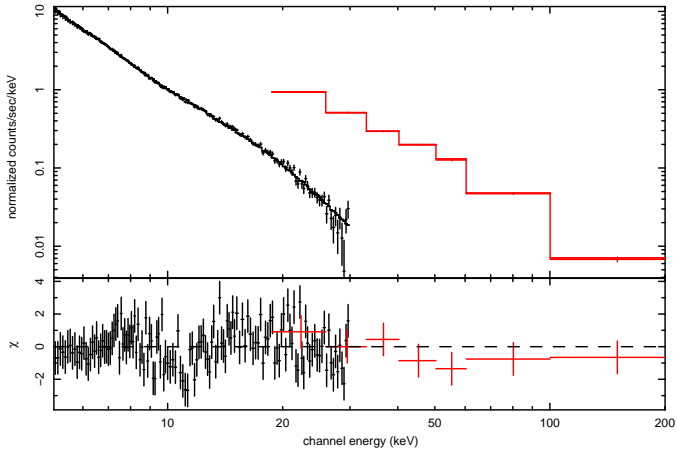}}
\caption{INTEGRAL JEM-X (black) and IBIS (red) spectra of XTE J1817-330, fit
with a disk model plus power-law.}
\label{fig_integral}
\end{figure}

\subsubsection{INTEGRAL observation}

We fit simultaneously JEM-X (6--30~keV) and IBIS/ISGRI (20--200~keV) spectral data 
obtained on 15--18 February 2006 with a two component model, \emph{diskbb + pow} (Figure~\ref{fig_integral}).
We add a free relative normalization constant to IBIS data 
to account for differences between the JEM-X and IBIS calibration. 
The best fit ($\chi^2_\nu=1.3$) is obtained with $kT_{\rm{in}}=0.95(\pm0.04)\,\mbox{keV}$, 
$K_{\rm {diskbb}}=\left( \frac{R_{in}/km}{D/10kpc}\right)^2 \cos i=1500\pm400$, 
photon index $\Gamma=2.64\pm0.04$ and power-law normalization 
$K=4.2\pm0.4\,\mbox{photons}\,\mbox{keV}^{-1}\,\mbox{cm}^{-2}\,\mbox{s}^{-1}$@1~keV. 
The relative normalization constant added to IBIS is 1.3. 
Other spectral models do not improve the fit. 
No cut-off is found in the spectrum up to 150~keV, and a power-law with a 
photon index 2.66$\pm$0.02 provides a good fit to the IBIS data alone ($\chi^2_{\nu}=0.92$). 
The observed flux is $3.90(\pm0.05) \times 10^{-9}\,\mbox{erg}\,\mbox{cm}^{-2}\,\mbox{s}^{-1}$ in 
JEM-X (5--30~keV) and  $1.55(\pm0.03)\times 10^{-9}\,\mbox{erg}\,\mbox{cm}^{-2}\,\mbox{s}^{-1}$ in 
IBIS (20--200~keV) (errors at 90\% confidence).

\begin{figure}
\centering
\resizebox{\hsize}{!}{\includegraphics[angle=0]{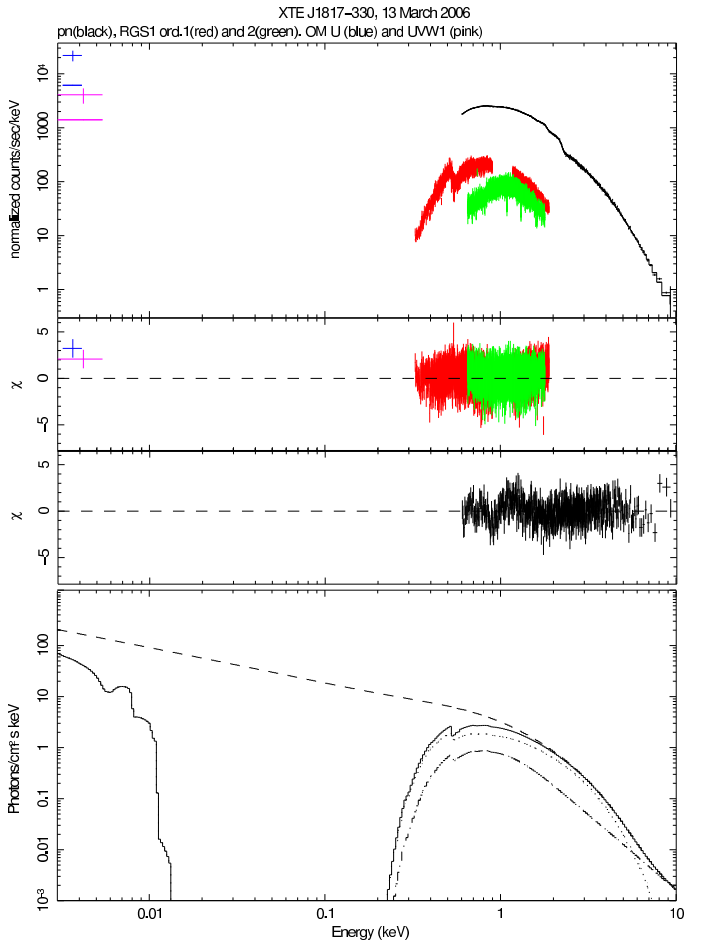}}
\caption{
\textbf{Upper panel:} XMM-Newton EPIC-pn (black), RGS1 first (red) and second (green) order spectra, 
and optical monitor data with filters U (blue) and UVW1 (purple) of XTE J1817-330, fitted with an 
absorbed accretion disk blackbody model (\emph{diskpn}) plus comptonization model (\emph{compTT}). 
\textbf{Second panel:} Residuals for RGS and OM. 
\textbf{Third panel:} Residuals for EPIC-pn.
\textbf{Lower panel:} Best-fit model (solid line), showing the
contributions of the disk (dotted line) and the \emph{compTT} (dash-dot line) models. 
The total model without absorption is plotted in dashed line, to show the contribution of the disk 
to the optical and UV bands. }
\label{fig_xmm}
\end{figure}

\begin{table*}
\centering
\caption{\label{tabpow} Models for simultaneous fit of EPIC-pn (0.6-10.0~keV), 
RGS1 (order 1 and 2, 0.3-2.0 keV) and OM (U and UVW1 filters) data 
(best-fit parameters with 3$\sigma$ confidence errors).}
\begin{tabular}{l c c}
\hline\hline
\noalign{\smallskip}
  \multicolumn{3}{c}{TBabs (diskbb + pow)}  \\
\noalign{\smallskip}
\hline
\noalign{\smallskip}
$N_{\rm H}\,(\mbox{cm}^{-2})$ 		&  	$2.3(\pm0.1)~\times 10^{21}$			&  $1.5\times 10^{21}$ (frozen)\\
\noalign{\smallskip}
$kT_{\rm{in}}\,(\mbox{keV})$		&	$0.70(\pm0.01)$ 				&  $0.70(\pm0.01)$	\\
$K^b_{\rm{diskbb}}$ 			&	$1900\pm50$ 					&  $2100\pm50$\\
\noalign{\smallskip}
Photon~index, $\Gamma$			&	$2.6\pm0.1$ 					& $1.6\pm0.10$	 \\
$K_{\rm{pow}}\,(\mbox{photons}\,\mbox{keV}^{-1}\,\mbox{cm}^{-2}\,\mbox{s}^{-1}$@1~keV)	& $230\pm40$	& $8\pm3$	 \\	
\noalign{\smallskip}
$\chi^2_{\nu}$(d.o.f)		&1.25 (6029) 		&	1.45 (6030) \\
\hline
\noalign{\smallskip}
\end{tabular}
\begin{list}{}{}
\item[$^{\mathrm{a}}$] All model fits include the oxygen absorption edge and lines 
listed in Table~\ref{tabox}.
\item[$^{\mathrm{b}}$] Normalization constant of the \emph{diskbb} accretion disk model, 
$K_{\rm{diskbb}}=\left( \frac{R_{in}/km}{D/10kpc}\right)^2 \cos i$,  where $R_{in}$ 
is the innermost disk radius, $D$ the distance to the source, and $i$ the inclination of the disk.
\end{list}
\end{table*}

\begin{table*}
\centering
\caption{\label{tabcomptt} Models for simultaneous fit of EPIC-pn (0.6-10.0~keV), 
RGS1 (order 1 and 2, 0.3-2.0 keV) and OM (U and UVW1 filters) data 
(best-fit parameters with 3$\sigma$ confidence errors).}
\begin{tabular}{l c c}
\hline\hline
\noalign{\smallskip}
  & TBabs (diskbb + compTT) & TBabs (diskpn + compTT) \\
\noalign{\smallskip}
\hline
\noalign{\smallskip}
$N_{\rm H}\,(\mbox{cm}^{-2})$ 		&  	$1.52(\pm0.05)~\times 10^{21}$	& $1.55(\pm$0.05)~$\times 10^{21}$\\
\noalign{\smallskip}
$kT_{\rm{in}}\,(\mbox{keV})$		&	$0.74(\pm0.01)$ 	&	$0.70(\pm0.01)$	\\
$K^b_{\rm{diskbb}}$ 			&	$1350\pm100$ & - \\
$R_{\rm{in}}$				&	 -			&	$6\rm{R}_{\rm g}$ (frozen)\\
$K^d_{\rm{diskpn}}$ 			&	-			&	$0.024\pm0.002$		\\
\noalign{\smallskip}
$kT_{\rm ph}\,(\mbox{keV})$		&	$0.20(\pm0.01)$ 	&	$0.21(\pm0.01)$ \\
$kT_{\rm e}\,(\mbox{keV})$		&	50 (frozen)		&	50 (frozen) \\	
$\tau$					&	$0.15(\pm0.02)$		&	$0.15(\pm0.02)$\\
$K^c_{\rm{compTT}}$ 			&	$0.040\pm0.003$		&	$0.038\pm0.003$\\
\noalign{\smallskip}
L$_{0.4-10\rm{keV}}\,(\mbox{erg}\,\mbox{s}^{-1})$	&	$1.2(\pm0.1)\times10^{38}({\rm D}/10\rm{kpc})^2$ &  $1.2(\pm0.1)\times10^{38}({\rm D}/10\rm{kpc})^2$\\
$\chi^2_{\nu}$(d.o.f)			&1.19 (6028) 		&	1.18 (6028) \\
\hline
\noalign{\smallskip}
\end{tabular}
\begin{list}{}{}
\item[$^{\mathrm{a}}$] All model fits include the oxygen absorption edge and lines 
listed in Table~\ref{tabox}.
\item[$^{\mathrm{b}}$] Normalization constant of the \emph{diskbb} accretion disk model, 
$K_{\rm{diskbb}}=\left( \frac{R_{in}/km}{D/10kpc}\right)^2 \cos i$,  where $R_{in}$ 
is the innermost disk radius, $D$ the distance to the source, and $i$ the inclination of the disk.
\item[$^{\mathrm{c}}$] Normalization constant of the \emph{compTT} comptonization model.
\item[$^{\mathrm{d}}$] Normalization constant of the \emph{diskpn} accretion disk model, 
$K_{\rm{diskpn}}=\frac{M^2 \cos i}{D^2 \beta^4}$, 
where \textit{M} is the black-hole mass in solar units, \textit{i} is the inclination of the disk, 
\textit{D} is the distance to the system in kpc and $\beta$ is the color/effective temperature ratio.
\end{list}
\end{table*}

\subsubsection{XMM-Newton observation}

We fit simultaneously the data of all XMM-Newton instruments active during the 13 March 2006 observation, i.e., 
EPIC-pn (0.6--10.0~keV), RGS (0.3--2.0~keV) and OM (exposures with U and UWV1 filters) data (Fig.~\ref{fig_xmm}).
In all fits in this section, the absorption edge and lines described in section~\ref{sec_ism} 
are included to minimize the RGS residuals. Error ranges given in this section are 3$\sigma$ confidence.

We use the {\it TBabs} model for the foreground absorption (Wilms~et~al.~\cite{wil00}). 
We cannot, however, use the same absorption model for the OM data, because
X-ray absorption models are not defined for UV and optical wavelengths.
We use thus the {\it redden} model in {\it xspec} for the OM data, 
an IR/optical/UV extinction model from Cardelli et al. (\cite{car89})
with transmission set to unity for energies higher than 15 eV.
During the fit, we keep the extinction in {\it redden} for the OM data linked 
to the absorbing hydrogen column density of the {\it TBabs} model affecting the X-ray data, by using the relation between 
optical extinction and hydrogen column density $N_{\rm H}=5.9\times10^{21}E_{B-V}\,\mbox{cm}^{-2}$
(Zombeck \cite{zom07}).

We first attempt to fit the continuum spectrum with an absorbed multi-temperature
accretion disk blackbody model ({\it diskbb} in {\it xspec}), plus 
a power-law accounting for the contribution of the comptonized photons in the hot corona.
A reasonable fit ($\chi^2_{\nu}=1.25)$ is obtained only for an absorption column density 
$N_{\rm{H}}=2.3(\pm0.3)\times10^{21}\,\mbox{cm}^{-2}$,
with $kT_{\rm{in}}=0.70(\pm0.01)\,\mbox{keV}$, 
$K_{\rm {diskbb}}=\left( \frac{R_{in}/km}{D/10kpc}\right)^2 \cos i=1900\pm100$ 
and photon index $\Gamma=2.6\pm0.1$ (Table~\ref{tabpow}). 
This high absorption column is in disagreement with the low reddening observed 
towards the source (Torres et al.~\cite{tor06}) and the low absorption
pointed out from RXTE, Swift, Chandra and INTEGRAL observations
(Remillard et al.~\cite{rem06}, Miller et al.~\cite{mil06a}, Steeghs et al.~\cite{ste06}, 
Miller et al.~\cite{mil06b}, Goldoni et al.~\cite{gol06}).
Fixing the absorption column to the interstellar value ($N_{\rm{H}}=1.5\times10^{21}\,\mbox{cm}^{-2}$)
leads to a poorer fit ($\chi^2_{\nu}=1.5$, see Table~\ref{tabpow} for details) with large residuals for the EPIC-pn. 
These poor results of the \emph{diskbb + pow} model are driven by the fact that the 
power--law component not only accounts for the hard energy band spectrum, but 
dominates the emission model at low energies, below 1~keV. 
If the power-law component accounts for the disk photons
inverse-comptonized in the corona to higher energies, 
it is unphysical that they dominate the emission at energies lower 
than the maximum emission of the disk.

Better fits are obtained using the {\it compTT} model for the corona component 
(Titarchuk \cite{tit94}, see Table~\ref{tabcomptt}). The electron temperature $kT_{\rm e}$ is fixed to 
50~keV in all fits. We do not consider it necessary to tie the temperature of the seed photons to the 
disk temperature because the best-fit is always obtained with temperatures lower than the maximum temperature
of the accretion disk. 
The best fit for the \emph{diskbb+compTT} model (with $\chi^2_{\nu}=1.19$) is obtained for 
an absorption column density $N_{\rm{H}}=1.52(\pm0.05)\times10^{21}\,\mbox{cm}^{-2}$, 
a disk with $kT_{\rm{max}}=0.71(\pm0.01)\,\mbox{keV}$ and 
$K_{\rm {diskbb}}=1350\pm100$, and a comptonizing hot plasma with
seed photons temperature $kT_{\rm ph}=0.20(\pm0.01)\,\mbox{keV}$, and
optical depth $\tau=0.15(\pm0.02)$.
 
Table \ref{tabcomptt} also lists the results of the fit using the 
{\it diskpn} model also available in {\it xspec} (Gierli\'nski et al.~\cite{gie99}). 
The {\it diskpn} model is an extension of the {\it diskbb} model, which takes 
into account the torque-free inner-boundary condition for the accretion disk. 
The results are very similar with both disk models, but 
we prefer {\it diskpn} both because of the more precise treatment of the inner boundary
and for its parametrization. We fix the inner radius in the {\it diskpn} model to 6$R_g$.
We obtain the best fit for the {\it diskpn + compTT} model ($\chi^2_{\nu}=1.18$) with 
an absorption column density of $N_{\rm{H}}=1.55(\pm0.05)\times10^{21}\,\mbox{cm}^{-2}$, and
a disk with $kT_{\rm{in}}=0.70(\pm0.01)\,\mbox{keV}$ and 
$K_{\rm {diskpn}}=\frac{M^2 \cos i}{D^2 \beta^4}=0.024\pm0.002$,
where \textit{M} is the black-hole mass in solar units, \textit{i} is the inclination of the disk, 
\textit{D} is the distance to the system in kpc and $\beta$ is the color/effective temperature ratio. 
The small effective temperature suggests a non-rotating black-hole, and 
justifies thus the choice of $R_{in}$.
The best-fit parameters for the {\it compTT} model component are the same as for the {\it diskbb + compTT} case.

The total unabsorbed flux from the EPIC-pn spectrum (0.4-10~keV) is 
$1.02(\pm 0.08)\times10^{-8}\,\mbox{erg}\,\mbox{cm}^{-2}\,\mbox{s}^{-1}$, 
indicating an X-ray luminosity of the source at the time of the observation of 
$1.2(\pm0.1)\times10^{38}({D}/10\rm{kpc})^2\,\mbox{erg}\,\mbox{s}^{-1}$.

\begin{table*}[t]
\centering
\caption{\label{tabox}Oxygen absorption features in the RGS spectra (errors indicate 3$\sigma$ confidence range.)}
\begin{tabular}{c c c c c c}
\hline\hline
\noalign{\smallskip}
Identification & Measured & FWHM & Eq. width & F-test & ISM rest-frame\\
 & $\lambda\,(\rm \AA)$ & $\lambda\,(\rm \AA)$ & $\lambda\,(\rm \AA)$ & Probability & $\lambda\,(\rm \AA)$\\
\noalign{\smallskip}
\hline
\noalign{\smallskip}
O~I~K$\alpha$ 	& $23.52\pm0.02$ & $<0.09$	& $0.07\pm0.02$ &$2.2\times10^{-25}$ & 23.507\\
O~II~K$\alpha$ 	& $23.35\pm0.03$ & $<0.27$	& $0.04\pm0.02$ &$1.8\times10^{-07}$ & 23.350\\
O~III~K$\alpha$ & $23.13\pm0.09$ & 0.1(fixed) 	& $0.03\pm0.03$ &$6.3\times10^{-02}$ & 23.114\\
O~I~K$\beta$ 	& $22.91\pm0.03$ & $<0.25$ 	& $0.06\pm0.02$ &$2.8\times10^{-11}$ & 22.887\\
O~VII~K$\alpha$ & $21.60\pm0.06$ & $<0.80$	& $0.04\pm0.03$ &$1.6\times10^{-05}$ & 21.602\\
\noalign{\smallskip}
\hline
\noalign{\smallskip}
		&  Observed~$\lambda (\rm \AA$)	& \multicolumn{4}{c}{Depth $(\tau)$ } \\
\noalign{\smallskip}
O~K~edge 	& $22.85\pm0.03$	& \multicolumn{4}{c}{$0.73\pm0.06$}  \\
\hline
\noalign{\smallskip}
\end{tabular}
\begin{list}{}{}
\item[$^{\mathrm{a}}$] Average ISM rest-frame wavelength, resulting from the 
average of several detections from Juett~et~al.~(\cite{jue04}) for OI, OII and OIII, 
and Yao~\&~Wang~(\cite{yao06}) for OVII.
\end{list}
\end{table*}

\subsection{Interstellar absorption lines \label{sec_ism}}

\begin{figure}
\centering
\resizebox{\hsize}{!}{\includegraphics{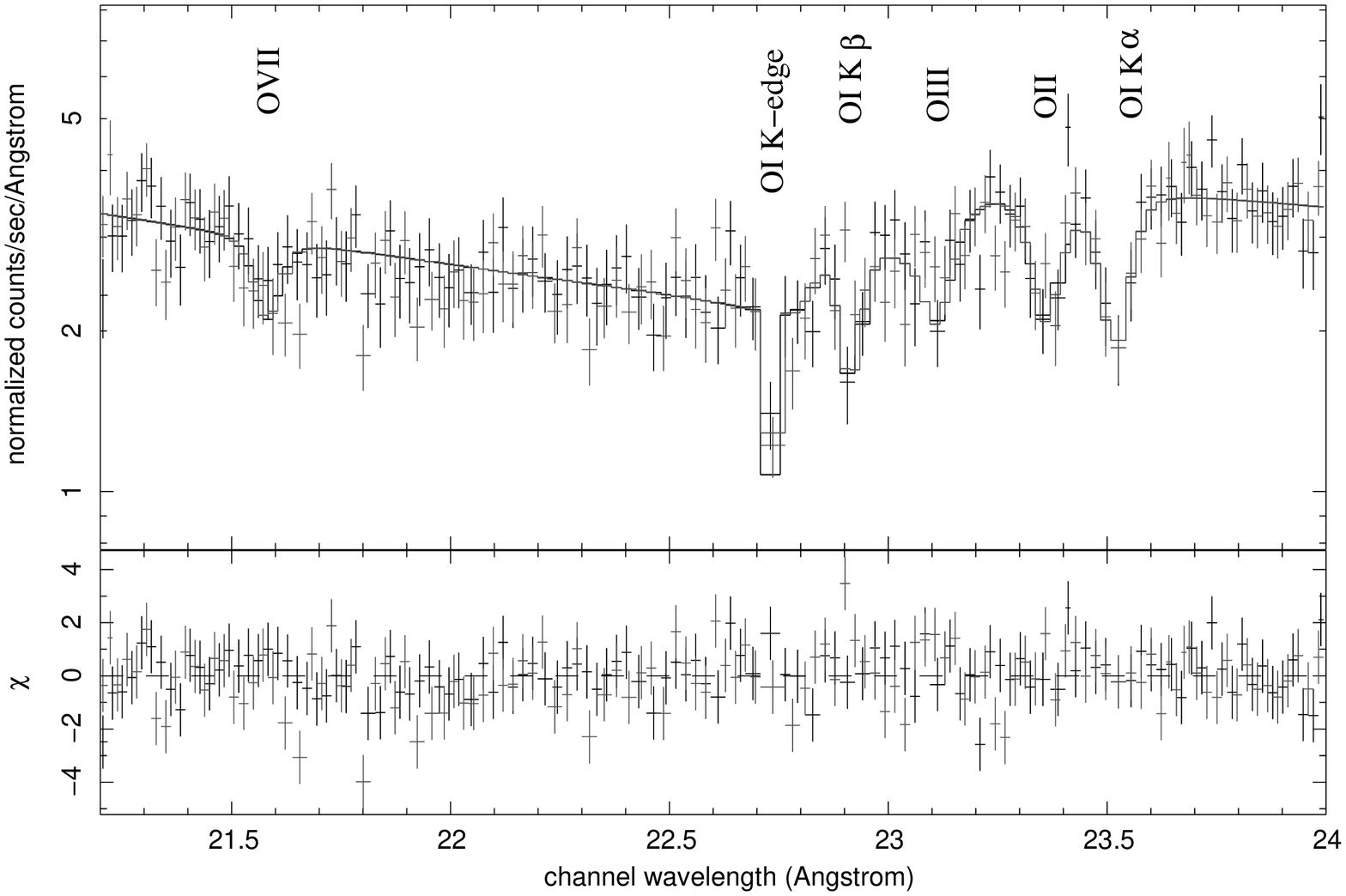}}
\caption{XMM-Newton RGS1 spectrum around the oxygen edge.}
\label{fig3}
\end{figure}

Several absorption features are detected in the RGS spectrum around the 
oxygen K edge (Fig.~\ref{fig3}). In order to explicitly fit the complex structure of the 
edge and surrounding lines, we have used the {\it vphabs} model in {\it xspec} as absorber with the oxygen 
abundance set to zero. We use Wilms et al.~(\cite{wil00}) solar abundances for other elements than oxygen, 
and Verner~et~al.~(\cite{ver96}) cross-sections.
We then fit the absorption features with an edge and five Gaussian lines (Table~\ref{tabox}).

The observed wavelengths are compatible with the rest frame energy 
of the lines. Taking into account the uncertainty in the 
wavelength determination for OI and OII lines (250~km/s for OI, 380~km/s for OII), 
we can discard the possible location of low ionized species in 
an intrinsic high-velocity wind of the source. Nevertheless, 
we can not safely distinguish between interstellar medium (ISM) absorption and 
some low velocity absorber intrinsic to the source for OIII or OVII
(with error ranges in the wavelength determination of 
1200~km/s for OIII and 800~km/s for OVII).
Interstellar oxygen lines around the oxygen K-edge
have previously been observed in the RGS spectra of other sources, 
and distinguished from the known instrumental components 
at 23.05 and 23.35 \AA (de Vries et al.~\cite{cor03}). 
Interstellar O~I, O~II and O~III absorption lines have also been detected
in the high resolution Chandra/HETGS spectra of several X-ray binaries 
and assigned to the ISM (Juett et al.~\cite{jue04}), and 
Yao~\&~Wang~(\cite{yao06}) used the absorption spectra of 4U~1820-303 
to determine ISM oxygen and neon abundances.
 

The O~I~K$\alpha$ line at 23.52~$\rm \AA$ was also detected in 
the RGS spectra of GRO~J1655-40 (Sala et al.~\cite{sal07}).
As in that case, we confirm the ISM origin of the line by comparing its equivalent width and 
the hydrogen column density to the source with that of other sources from the previously mentioned works,
with which it is consistent.


The RGS1 spectrum of XTE~J1817-330 also shows an absorption line at 21.6~$\rm \AA$, which corresponds to
O~VII~K$\alpha$. O~VII indicates the presence of a hot absorbing gas, at T$\sim10^6$~K, and can be
produced either by the hot ISM component, or by some intrinsic warm absorber.
O~VII was detected in the Chandra/HETGS spectrum of GX~339-4 (Miller et al.~\cite{mil04}). 
The authors discussed in that case the difficulties with ascribing the absorption lines to the hot ISM and argued 
that the origin of the absorption was an intrinsic warm absorber close to the black hole. 
O~VII lines have also been detected at red-shift zero in the spectra of 
several AGN, opening a debate on the location of the absorber: either in the intergalactic
medium of the Local Group, in the Galactic halo or in the Galactic ISM. 
More recently, not only has O~VII been identified as ISM absorption in the spectra of 
several Galactic low mass X-ray binaries, but it has even been used as tracer of the hot component of 
the Galactic ISM (Yao~\&~Wang~\cite{yao05,yao06}).

\begin{figure}
\centering
\resizebox{\hsize}{!}{\includegraphics{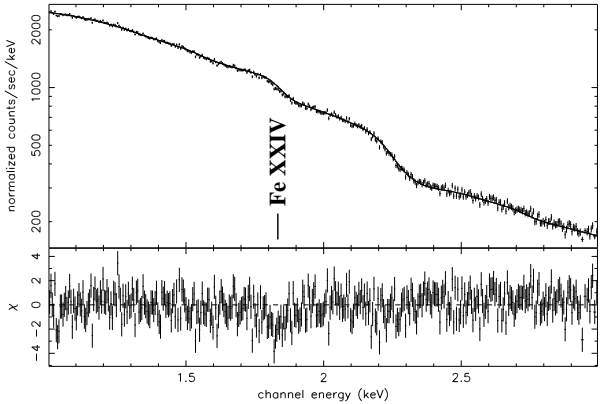}}
\caption{XMM-Newton EPIC-pn spectra in the energy range 1-3~keV, showing 
an absorption feature at $\sim$1.8~keV most probably due to interstellar Fe~XXIV.}
\label{fig5} 
\end{figure}

The residuals of the EPIC-pn spectrum show an absorption feature at 1.83($\pm0.03$)~keV 
($6.8\pm0.1\,\rm\AA$, Fig.~\ref{fig5}). 
This line lays in an energy range with rapidly changing effective area (close to Si edge), 
so its significance must be taken with care. 
We fit this line with a Gaussian with width fixed to 
the resolution of the EPIC-pn camera at this energy (100~eV), 
and obtain an equivalent width of $4\pm1$~eV ($0.015\pm0.004\,\rm \AA$, 90\% confidence error)
and an F-test probability of $2.4\times10^{-13}$.
A narrow absorption line near 1.82~keV has been also observed in the spectra of several AGN, and 
identified as the 1.821~keV line transition of Fe~XXIV (Haberl et al.~\cite{hab06}). The strongest line was found 
for MCG-6-30-15, which also showed another Fe~XXIV line, as well as lines 
from other Fe ions (Turner et al.~\cite{tur04}).
In all cases, the line was interpreted as ISM absorption. 
Supernova explosions or interaction of supernova remnants with the ISM
can heat the interstellar gas up to $\sim$2.5~keV, which fits with the temperature range of maximum 
emissivity of Fe~XXIV (1.5-2.5~keV).


\section{Discussion\label{sec_dis}}

The presented X-ray spectrum of XTE~J1817-330 in March 2006 shows the source in a typical
high/soft state for black hole transients. From the best-fit parameters of the continuum spectrum, 
some constrains are obtained for the distance to the source and the mass of the compact object.

Both the inner temperature and radius of the accretion disk point to the presence of 
a black hole in the system. The inner temperature of the disk (0.70$\pm$0.01~keV) is 
too low for a neutron star X-ray binary, with typical temperatures of 1.4-1.5~keV, while
for non-rotating black hole binaries it is typically less than $\sim$1.3~keV (Tanaka \cite{tan95}). 

\begin{figure}
\centering
\resizebox{\hsize}{!}{\includegraphics[angle=0]{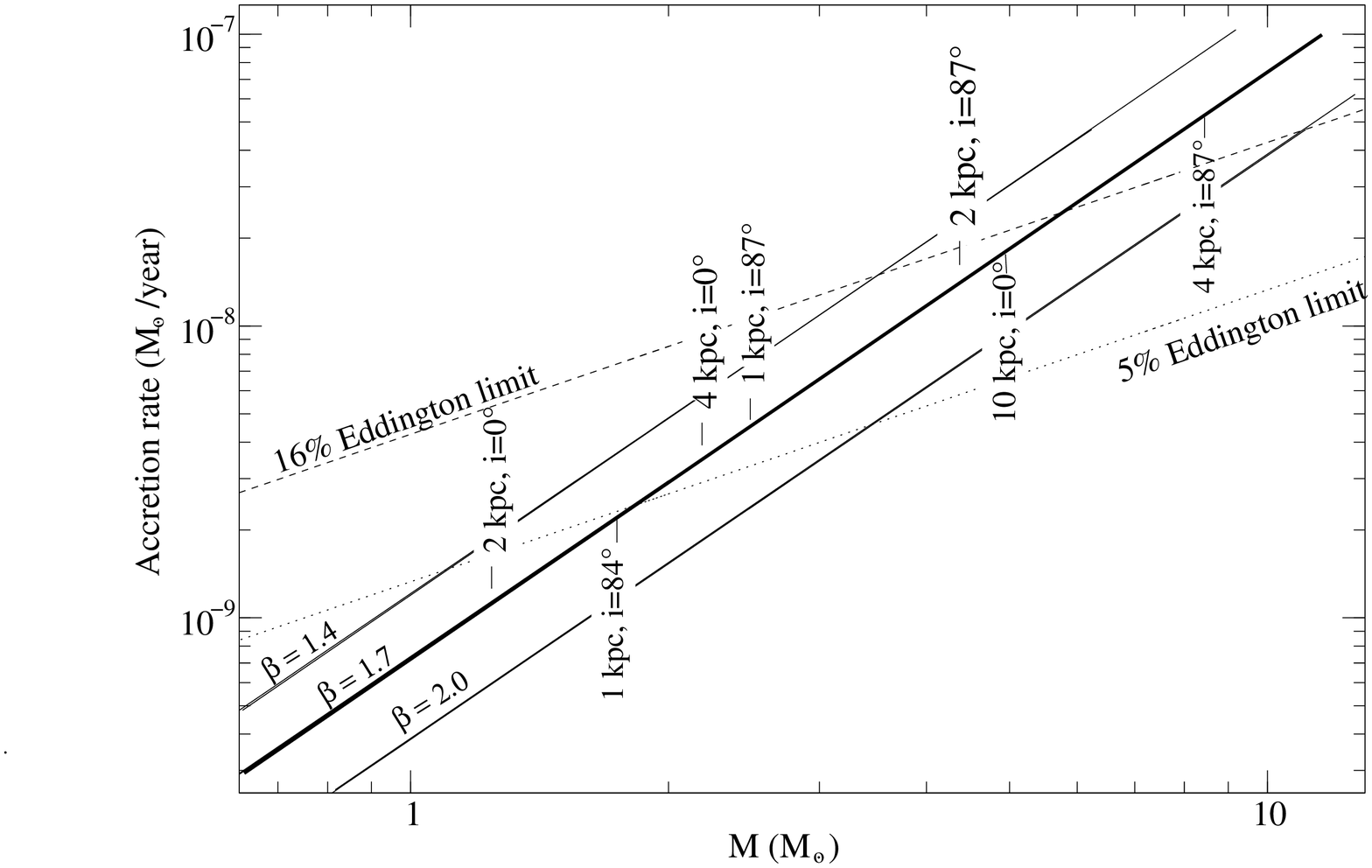}}
\caption{Accretion rate as a function of the black hole mass derived from the {\it dispkn} model, 
for different inclinations and distances. Over plotted are shown the accretion rate corresponding 
to 5\% and 16\% of the Eddington limit. The thick solid line is for $\beta=1.7$, while the thin
solid lines show the accretion rate for $\beta=1.4$ and $\beta=2.0$. 
The area between  5\% and 16\% of the Eddington limit show the possible values for the source 
at the time of the observation. This implies a maximum mass for the accreting object of 10~M$_\odot$.}
\label{figmm} 
\end{figure}

The normalization constant {\it K} of the {\it diskpn} model is 
related to the mass of the compact object {\it M}, the 
distance to the source {\it D} and the inclination of the disk {\it i} as $K=\frac{M^2 cos(i)}{D^2 \beta^4}$,
where $\beta$ is the color/effective temperature ratio. Furthermore, the accretion rate can be obtained
from the mass of the compact object and the maximum temperature of the disk (Gierli\'nski et al.~\cite{gie99}). 
Assuming $\beta=1.7$ and with the best fit value for the normalization of the {\it diskpn} model ($K=0.05$), 
we plot in Figure \ref{figmm} the accretion rate as a function of the mass of the compact object, 
giving different values for inclinations and distances (from 1 to 10~kpc).
At the time of the XMM-Newton observations, the flux of the source had decreased by a factor of 6 
with respect to the maximum registered by RXTE. 
Assuming that at the time of the maximum the accretion rate was 30\% of the Eddington limit,
$\rm M^{\rm{acc}}_{\rm{Edd}}$, we overplot in Figure~\ref{figmm} the accretion rate corresponding 
to 5\% of the Eddington limit. This sets a minimum mass for the central object of 2~M$_{\odot}$. 

To set an upper limit, we also overplot the accretion rate corresponding
to 16\% of $\rm M^{\rm{acc}}_{\rm{Edd}}$ (corresponding to the source being close to the Eddington 
limit at the time of the maximum).
This sets an upper limit for the mass for the central object of 6~M$_{\odot}$.

The above determined limits are obtained assuming $\beta=1.7$. In Figure~\ref{figmm} we also 
plot the accretion rate for $\beta=1.4$ and $\beta=2.0$ to establish some ``harder'' limits for 
the mass. This leads to a hard upper limit of 10.0~M$_\odot$ for the black hole in XTE~J1817-330.
Other sources of uncertainty in this upper limit, like the dependence of the disk parameters on
the hard component model, are much smaller and thus negligible in this exercise: 
as determined in section~\ref{sec_ana}, the disk temperature remains the same within 5\%,
independent of the hard component model; the disk normalization parameter experiences 
more differences depending on the hard component model and the absorption, changing up to a factor 
1.5. However, any of these uncertainties are reflected in Figure~\ref{figmm} by a change in 
the accretion rate of the order of the line width in the plot, completely negligible in 
the rough determination of mass limits presented here.


The absorption column obtained with the  \emph{diskbb+compTT} and \emph{diskpn+compTT} model fits
is in agreement with the low absorption already pointed out at the 
discovery of the source with RXTE (Remillard et al.~\cite{rem06}), 
and confirmed later on in UV and X-rays with pointed 
observations of RXTE (Miller et al.~\cite{mil06a}), Swift/UVOT (Steeghs et al.~\cite{ste06}), 
Chandra (Miller et al.~\cite{mil06b}) and INTEGRAL (Goldoni et al.~\cite{gol06}).
The low absorption towards XTE~J1817-330 makes this new black hole candidate
an ideal target for optical observations in the quiescent state, which could provide the 
dynamical confirmation of the presence of a black hole in the system.

The absorbing hydrogen column is also compatible within the errors with the average galactic column density 
in the source direction, 1.57$\times10^{21}\rm{cm}^{-2}$ (Dickey \& Lockman~\cite{dic90}). This locates
XTE~J1817-330 behind the galactic slab of interstellar medium and 
provides a lower limit for the distance of 1~kpc.

From Figure \ref{figmm}, taking the 6~M$_{\odot}$ upper limit for the central object mass and assuming 
a face on disk, we obtain an upper limit for the distance of 10~kpc. 

The type of the secondary star can be constrained from the non-detection of 
the source previous to outburst in the DSS (V$>$22~mag). 
With the extinction towards the source derived from the N$_{\rm H}$, A$_{\rm V}=0.76$, 
the lower limit on the absolute magnitude is M$_{\rm V}> 6$~mag (for the maximum possible
distance, 10~kpc). 
A limit on the absolute magnitude of M$_{\rm V}> 6$~mag implies 
that the secondary star must be a K-M star, and a giant can be excluded, 
even for a 10~kpc distance. 
We therefore conclude that XTE~J1817-330 is a low-mass X-ray binary.

From Figure~\ref{fig_xmm} it is clear that the accretion disk model underestimates
the U and UV emission of XTE~J1817-330.  Rykoff~et~al.~(\cite{ryk07}) find the same with the optical monitor on 
board Swift. This indicates that the U and UV emission is not dominated by the viscous dissipation 
of the disk. Rykoff~et~al.~(\cite{ryk07}) find, in addition, that the e-folding time of the NUV light curve is
roughly twice that of the hard X-ray light curve, and they conclude that this is consistent 
with the optical and UV emission being dominated by reprocessed hard X-ray emission.

\begin{acknowledgements}
XMM-Newton and INTEGRAL projects are ESA Science Missions with instruments
and contributions directly funded by ESA Member States and the
USA (NASA). Both are supported by BMWI/DLR (FKZ 50 OX 0001), the Max-Planck
Society and the Heidenhain-Stiftung.
We acknowledge the RXTE/ASM team for making quick-look results available for public use. 
GS acknowledges a postdoctoral fellowship of the Spanish Ministerio de Educaci\'on y Ciencia.
EB is supported through DLR (FKZ 50 OR 0405).
\end{acknowledgements}

\end{document}